\documentclass[a4paper]{article}
\usepackage{epsfig,amsfonts,amsmath,amssymb,setspace,multicol,textcomp,xcolor}
\usepackage[numbers]{natbib}
\usepackage{longtable}
\usepackage[T1]{fontenc}
\makeatletter
\renewcommand{\@oddhead}{\hfil \textit{I.\,Sushch for the H.E.S.S. Collaboration}}
\renewcommand{\@evenfoot}{\hfil \thepage \hfil}
\renewcommand{\@oddfoot}{\hfil \thepage \hfil}

\newenvironment{figurehere} {\def\@captype{figure}} {}

\makeatother

\renewenvironment{thebibliography}[1]{\begin{oldthebibliography}{#1}\setlength{\parskip}{0ex}\setlength{\itemsep}{0ex}}{\end{oldthebibliography}}
\newcommand{\myHeder}[2]{\raisebox{+2.8cm}[.2cm][.2cm]{\vbox{\hbox to\textwidth{ #2 \hfill}\hbox to\textwidth{ #1 \hfill}}}}
\newcommand{\myCopy}[2]{\raisebox{-20.7cm}[.2cm][.2cm]{\vbox{\hbox to\textwidth{ #2 \hfill}\hbox to\textwidth{ #1 \hfill}}}}

\def\hess{H.E.S.S.}

\def\fermi{\textit{Fermi}-LAT}

\usepackage[normalem]{ulem}

\begin{document}
\fontsize{11}{11} \selectfont
\title{\bf The Galactic sky through H.E.S.S. eyes}
\author{\textsl{I.~Sushch$^{1,2}$\thanks{\tt iurii.sushch@nwu.ac.za}, for the H.E.S.S. Collaboration}\vspace*{-1ex}}
\date{\vspace*{-13ex}}
\maketitle
\myHeder{\hspace*{-5ex} \fontsize{8}{8}\selectfont }{}
\myCopy{\fontsize{8}{8}\selectfont\sl \vspace*{-0.1ex} }{}
\begin{center} 
{\small  $^{1}$Centre for Space Research, North-West University, 11 Hoffman Street, 2531, Potchefstroom, South Africa\\
        $^{2}$Astronomical Observatory of Ivan Franko National University of L'viv, vul. Kyryla i Methodia, L'viv, Ukraine\\}
\end{center}
\vspace*{-4ex}
\begin{abstract}
The High Energy Stereoscopic System (H.E.S.S.) is an array of five imaging atmospheric Cherenkov telescopes. Since 2003 it has been operating in the configuration of four 12\,m telescopes complemented in 2012 by a much bigger 28\,m telescope in the centre of the array. It is designed to detect very high energy (VHE) gamma-rays in the range of $\sim 20$\,GeV to $\sim50$\,TeV. Over the past decade it performed extremely successful observations of the Galactic plane, which led to the discovery of about 70 sources amongst which the most numerous classes are pulsar wind nebulae, supernova remnants and binary systems. Recently \hess\ also discovered the VHE emission from the Vela pulsar, which became the second pulsar detected at TeV energies after the Crab pulsar. An overview of the main \hess\ discoveries in our Galaxy and their implications on the understanding of physical processes is discussed in this paper. \\[1ex]
\end{abstract}
\begin{multicols}{2}

\section*{\sc introduction}
\vspace*{-1ex}
\indent\indent Over the past decade, very high energy (VHE; $E>100$\,GeV) gamma-ray astronomy has become one of the most popular and fast-developing branches of the observational science and a driver of theoretical models in several topical areas of modern astrophysics and cosmology. The current generation of imaging atmospheric Cherenkov telescopes (\hess, VERITAS and  MAGIC) represent a breakthrough, opening up a window to the previously largely unexplored VHE Universe and its mysteries. An incredibly successful period of operation of these ground-based instruments resulted in detection of more than 100 VHE gamma-ray sources\footnote{For the current status of the population of the VHE gamma-ray sources check {\sl TeVCat}, an online TeV gamma-ray catalogue, at {\tt http://tevcat.uchicago.edu}}. This remarkable scientific breakthrough would not be possible without the High Energy Stereoscopic System (H.E.S.S.), which played a major role in the opening of the field of gamma-ray astronomy, becoming the main instrument in the southern hemisphere. 

\hess\ is an array of five imaging atmospheric Cherenkov telescopes located  in  the  Khomas  Highland  of  Namibia at an altitude of 1800\,m above sea level~\cite{2006A&A...457..899A}. During the summer of 2012 the array of the first four 12\,m telescopes was completed with the addition of a much larger 28\,m telescope in the centre of the array. This upgrade expanded the energy coverage of the instrument down to $\sim 20$\,GeV and increased the system's sensitivity. 

\begin{figure}[!t]
\centering
\begin{minipage}{.98\linewidth}
\centering
 \epsfig{file=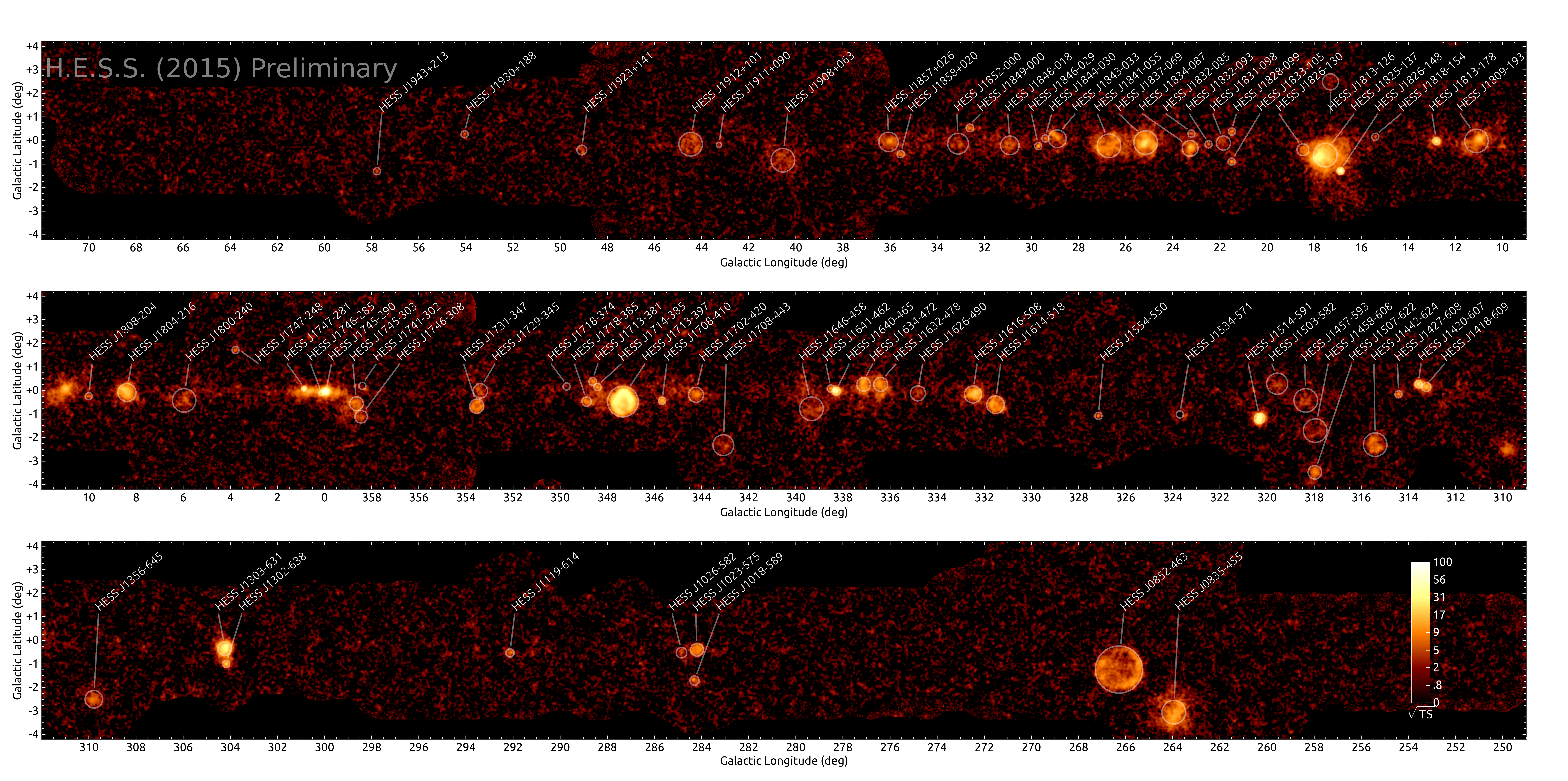,width=\linewidth}
\caption{The TS map of the \hess\ Galactic Plane Survey.}\label{hgps_fig}
\end{minipage}
\end{figure}

For the epoch of the \hess\,I observations (four telescope array), the sources in our Galaxy can be summarised in the \hess\ 
Galactic Plane Survey (HGPS; Fig.\,\ref{hgps_fig}) combining the data collected during the period starting from 2004 to 2013~\cite{HGPS}. The total of roughly 2800 hours of high quality observations in the Galactic longitude range of $250^{\circ}$ to $65^{\circ}$ and Galactic latitude range $\lvert b \rvert < 3.5^{\circ}$ are included in the survey. The HGPS reveals the diverse population of cosmic accelerators in the Galaxy resulting in the catalogue of 77 VHE sources. This includes 13 complex sources (Supernova Remnants, SNRs, and Galactic centre region) which were excluded from the analysis pipeline. Only sources with $\mathrm{TS}>25$ were included in the catalogue\footnote{Test Statistics -- the likelihood ratio of a model with the additional source at a specified location and a model without the additional source}. The data analysis was performed for the energy range of $0.2-100$\,TeV. The catalogue comprises 12 pulsar wind nebulae (PWNe), 6 SNRs, 6 composite objects and 3 binary systems (Fig.\,\ref{class_pie}). Fifty sources remain unidentified, mainly due to multiple associations, but also due to the lack of counterparts at other wavebands. Five new sources were discovered in the HGPS: HESS\,J1813$-$126, HESS\,J1826$-$130, HESS\,J1828$-$099, HESS\,J1832$-$085, and HESS\,J1844$-$030. Most of these sources were not detected before due to their proximity to the other, more extended \hess\ sources, and only a highly increased amount of data allowed one to discriminate them from their companions. Some of these new sources are coincident with the known pulsars, which suggests that they might be PWNe. HESS\,J1844-030 is coincident with the catalogued SNR~G29.4+0.1.
\end{multicols}

\begin{multicols}{2}
Beyond our Galaxy, more than 30 sources were discovered and associated with active galactic nuclei (AGNi). This population is dominated by the blazars of the BL Lacertae type.

\begin{figurehere}
\centering
\begin{minipage}{.98\linewidth}
\centering
\epsfig{file=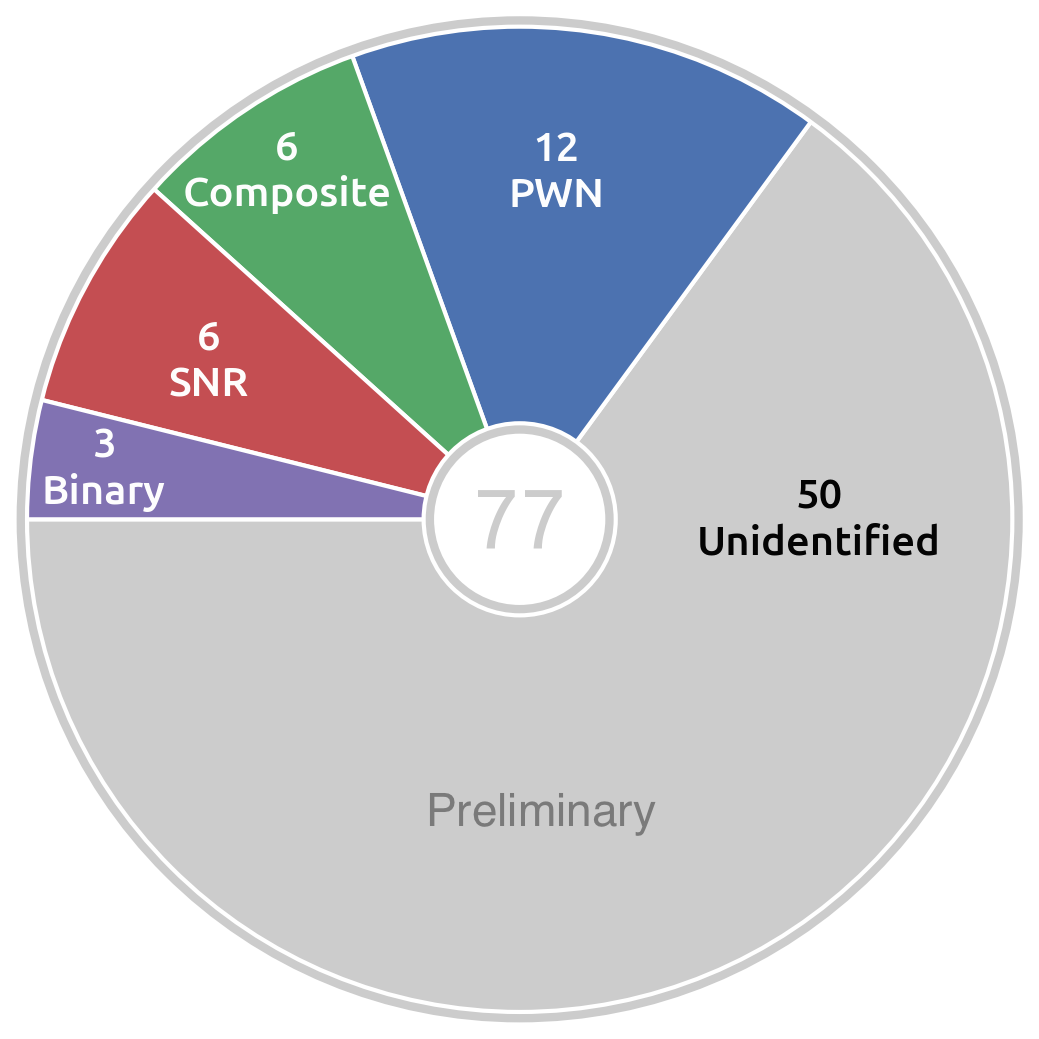,width=\linewidth}
\caption{Source classification of the HGPS.}\label{class_pie}
\end{minipage}
\end{figurehere}
\vspace*{1ex}

The observational strategy of \hess\ has been evolving over the years. During the first years of operation, H.E.S.S. was opening a new field of gamma-ray astronomy, discovering many new sources. Each new detection was treated as a major discovery, but in many cases limited exposure did not allow deep studies of the spectrum and morphology of the source. However, in recent years with the continuously growing population of gamma-ray sources, scientific priorities had shifted towards the better understanding of the nature of the gamma-ray emission from these sources. This led to deeper observations of specific objects in attempt to reveal and explain the physical processes generating VHE emission. The detection of numerous sources belonging to one class of objects (PWNe, SNRs, AGNi) allowed for population studies, which led to the investigation of common properties of sources of the same class.  

The goal of this paper is to give an overview of the current status of the Galactic sky as seen with H.E.S.S. with an emphasis on the recent results obtained during the last few years. Note that some of the results discussed here are preliminary results which were presented for the first time at the 34th Internation Cosmic Ray Conference in summer of 2015. For the most recent review of the \hess\ Galactic sky please consult~\cite{2013AdSpR..51..258D}, and for the most recent reviews of the VHE gamma-ray astronomy see, e.\,g.,~\cite{2014BrJPh..44..450H,2013FrPhy...8..714R}.  

\vspace*{-3ex}
\section*{\sc supernova remnants}
\vspace*{-1ex}
\indent\indent Supernova remnants are the remains of the supernova explosions of massive stars at the end of their evolution. As a result of this explosion the outer layers of the star are blown off into the surrounding medium, heating it up. The expansion of the SNR into the medium creates a shock wave at which particles (electrons and protons) can be accelerated to extremely high energies. The theory of diffusive shock acceleration at shock fronts~\cite{1983RPPh...46..973D} predicts the generation of accelerated particle populations in SNRs which, interacting in turn with ambient photon fields (electrons) or ambient matter (protons), can produce VHE gamma rays (see e.\,g.~\cite{1970RvMP...42..237B}).

\begin{figurehere}
\centering
\epsfig{file=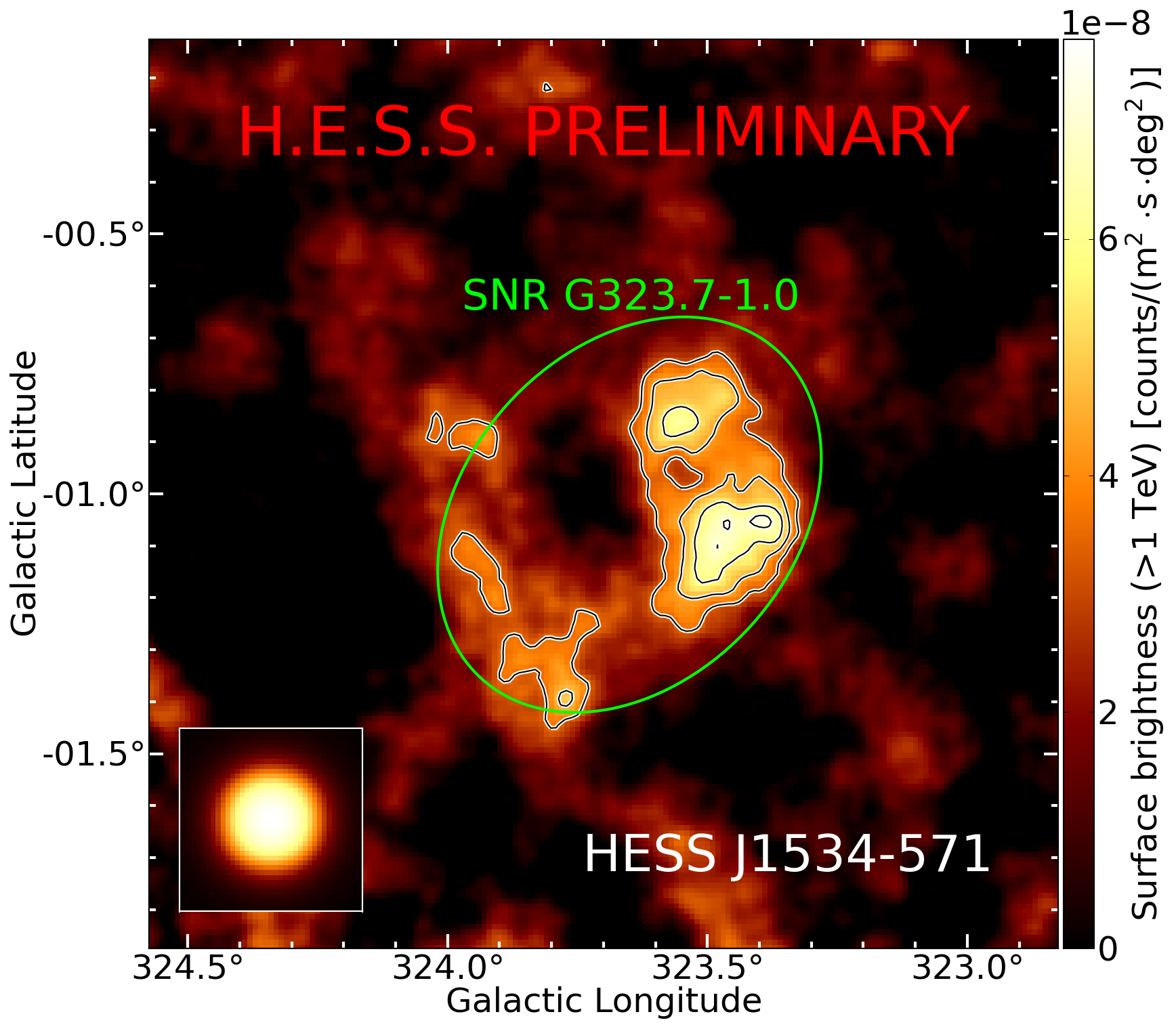,width=\linewidth}
\caption{The surface brightness map of HESS\,J1534$-$571. The green ellipse indicate the position of the radio SNR G323.7$-$1.0. The figure is taken from~\cite{HGPS}.}\label{j1534}
\end{figurehere}
\vspace*{1ex}

SNRs are the second most numerous class of VHE gamma-ray Galactic sources. Approximately 50 gamma-ray sources detected in the \hess\ Galactic Plane Survey (HGPS) are spatially coincident with SNRs detected in radio and higher frequency observations. However, the VHE emission can firmly be associated with SNRs for only 7 \hess\ sources: RX\,J0852-4622 (Vela Jr.) 
\cite{2005A&A...437L...7A, 2007ApJ...661..236A}, RX\,J1713.7-3946~\cite{2004Natur.432...75A, 2006A&A...449..223A, 2007A&A...464..235A,2002Natur.416..823E}, RCW\,86~\cite{2009ApJ...692.1500A},  SN\,1006\footnote{Located outside the Galactic Plane and, thus, not included in the HGPS}~\cite{2010A&A...516A..62A}, G323.7-01.0\footnote{HESS\,J1534$-$571 -- a new source detected in the HGPS}~\cite{HGPS}, G353.6$-$0.7~\cite{2011A&A...531A..81H}, and W\,28~\cite{2008A&A...481..401A}. All except the last one are shell-type SNRs with a resolved shell-like TeV morphology. For RCW\,86 the TeV shell was resolved only recently in a detailed morphology study which benefited from significantly improved statistics compared with the discovery paper~\cite{RCW86_proc}. G353.6$-$0.7 (or HESS\,J1731$-$347) is the first SNR discovered serendipitously in VHE gamma-rays and only later confirmed by radio and X-ray observations~\cite{2008ApJ...679L..85T, 2010ApJ...712..790T}. A new TeV shell-like source HESS\,J1534$-$571 (Fig.\,\ref{j1534}) was detected recently in the HGPS~\cite{HGPS} coincident with the radio SNR~G323.7-1.0 and thus firmly identified as SNR~\cite{2015arXiv150903872P}. Moreover, there are several other SNR candidates (with resolved TeV shell-like morphology) with the most prominent example being HESS\,J1912$+$101~\cite{2008A&A...484..435A} (Fig.\,\ref{j1912}). These, however, cannot be firmly identified as SNRs due to the lack of SNR counterparts at other wavelengths~\cite{2015arXiv150903872P}. Unfortunately, the TeV data alone are unable to firmly identify the shell-like source as an SNR as there are other astrophysical objects that potentially may appear shell-like while being potential TeV gamma-ray emitters, such as superbubbles or wind-blown cavities into which hadronic particles are diffusing~\cite{2015arXiv150903872P}.

\begin{figurehere}
\centering
\epsfig{file=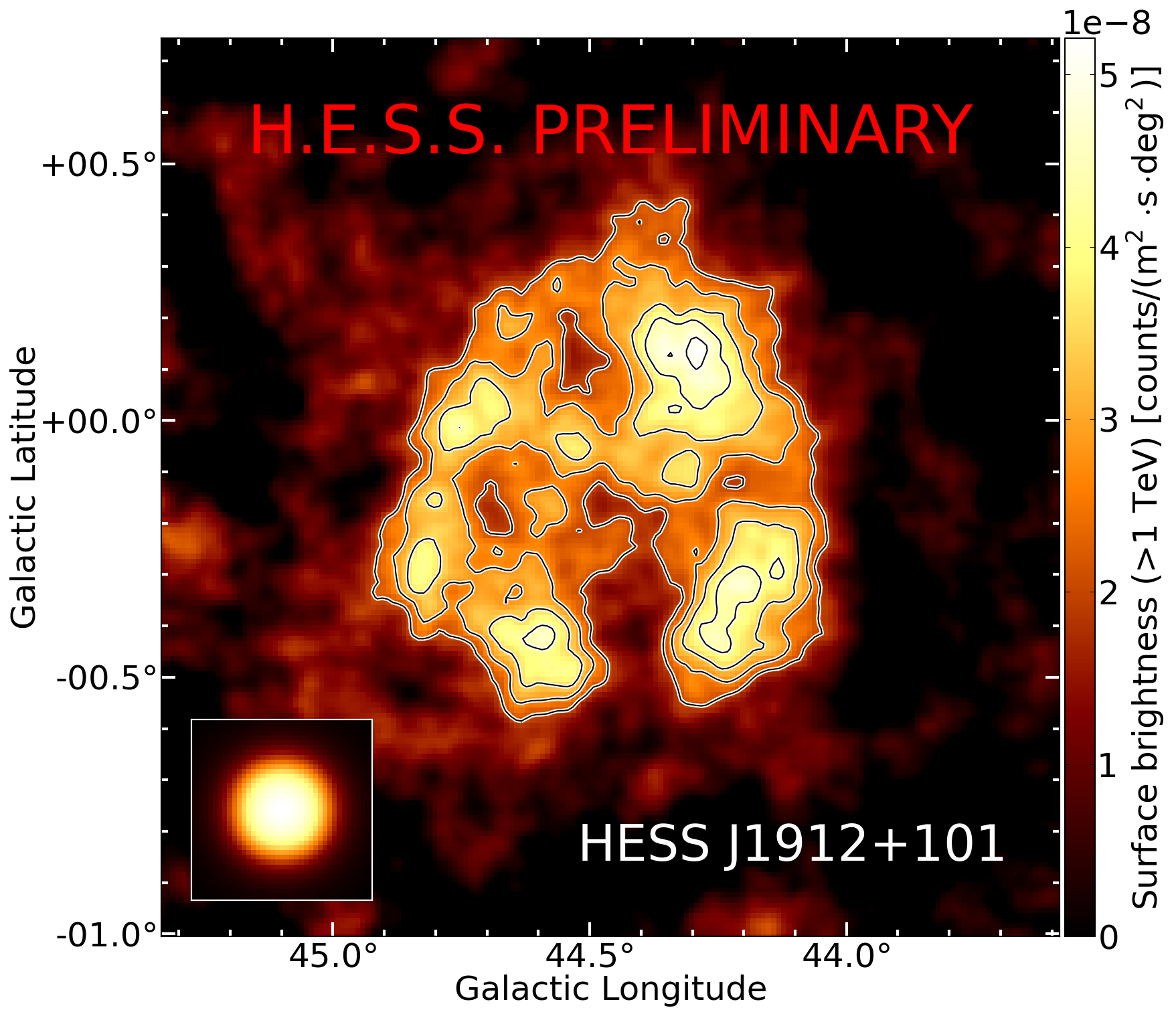,width=\linewidth}
\caption{The surface brightness map of HESS\,J1912$+$101. The figure is taken from~\cite{HGPS}.}\label{j1912}
\end{figurehere}
\vspace*{1ex}
 
The list of shell-type SNRs detected at TeV energies can be completed by the sources detected in the northern sky by VERITAS and MAGIC --- Cas\,A~\cite{2010ApJ...714..163A,2001A&A...370..112A, 2007A&A...474..937A}, Tycho~\cite{2011ApJ...730L..20A}, and IC\,443~\cite{2009ApJ...698L.133A,2007ApJ...664L..87A}. However, only for IC\,443 could the TeV shell recently be resolved~\cite{IC443_humensky}.

The most important question, which studies of the VHE emission from SNRs are expected to be able to answer, pertains to the origin of the Galactic cosmic rays. Galactic cosmic rays are believed to be mainly produced at the shocks of SNRs via acceleration of protons and electrons. When accelerated to very high energies, electrons and protons can in turn generate VHE gamma-rays via inverse Compton scattering on ambient photon fields and the bremsstrahlung process (electrons) and proton-proton interactions (protons). Cosmic rays consist of  99\,\% protons, thus any evidence of hadronic nature of the detected gamma-ray emission from SNRs can be treated as an indirect confirmation of the hypothesis that the Galactic cosmic rays originate in SNRs.

The spectral shape of the most of TeV SNRs can be described with both leptonic and hadronic scenarios with a slight preference for the leptonic one. This is not quite surprising as, usually SNRs expand into a rather rarefied medium created by their progenitor stars. However, for several examples in cases where SNRs interact with molecular clouds with much higher matter density, the hadronic scenario is much more preferable. This appear to be the case for such sources as e.\,g. IC\,443~\cite{2009ApJ...698L.133A, 2013Sci...339..807A,2007ApJ...664L..87A}, \hess\ SNR W\,28~\cite{2008A&A...481..401A}, and the GeV SNR detected by \fermi\ W\,44~\cite{2013Sci...339..807A}. Usually these are middle-aged SNRs which feature an escape of high-energy particles which then interact with molecular clouds producing VHE gamma-rays. The spatial offset of the gamma-ray emission region compared to the emission region at lower energies provides evidence for particle escape. Such detections of the 
SNRs interacting with molecular clouds are the first direct indications of the effective proton acceleration at SNR shocks.  

Detection of TeV SNRs (or SNR candidates) which do not have counterparts at X-ray energies (such as HESS\,J1912$+$101) becomes an important method to trace hadronic dominated SNRs. Lack of non-thermal X-ray emission suggests insufficient amounts of high energy electrons, and thus the gamma-ray emission from such sources can hardly be explained in the leptonic scenario.

\begin{figurehere}
\centering
\epsfig{file=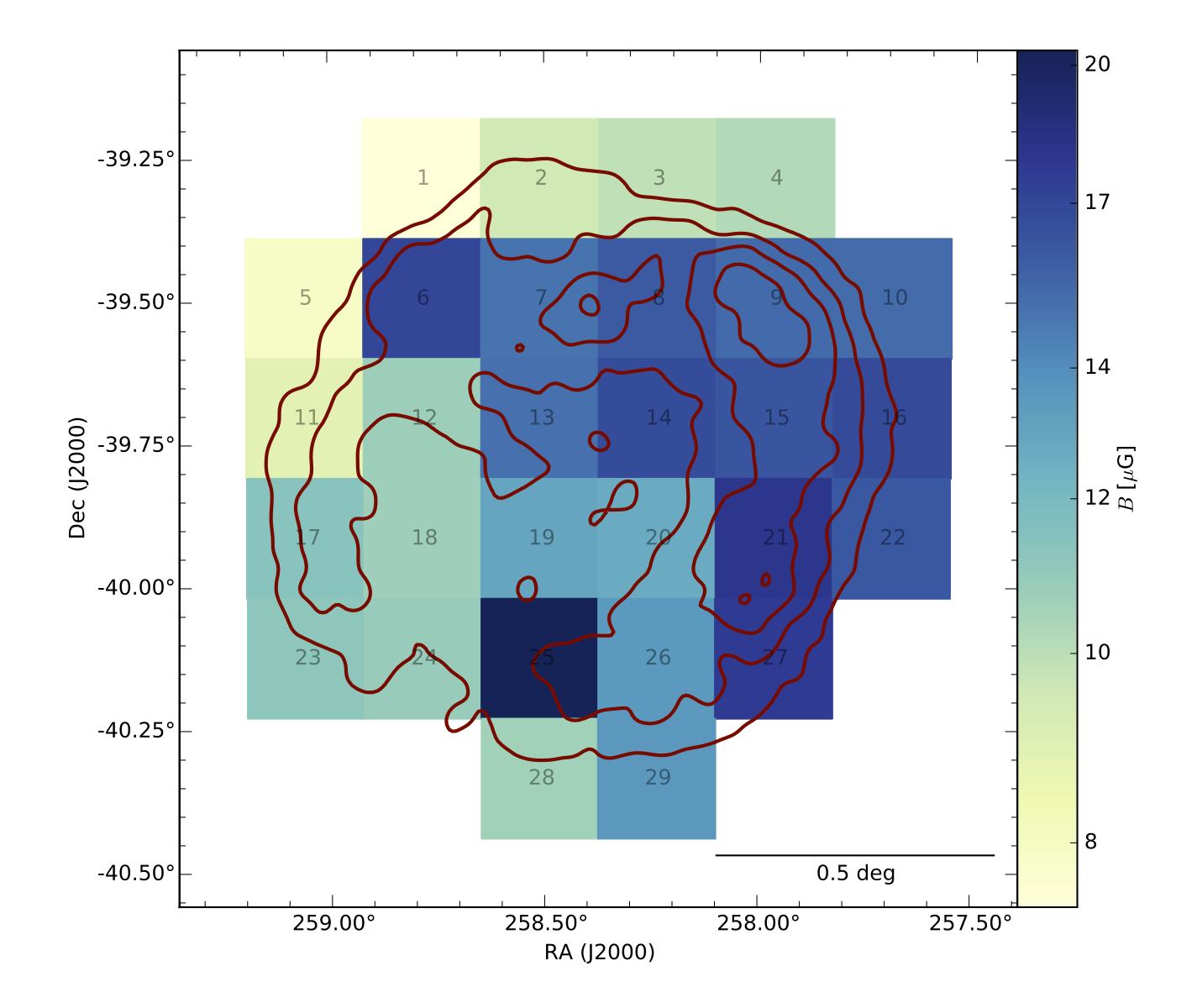,width=\linewidth}
\caption{The magnetic field map of RX\,J1713.7-3946. The figure is taken from~\cite{j1713_proc}.}\label{j1713_Bmap}
\end{figurehere}
\vspace*{1ex}

The focus of \hess\ observations on specific sources led to very detailed studies of a number of objects including some SNRs. One of the most remarkable examples is RX\,J1713.7-3946. New measurements based on $\sim150$\,h of observations benefit from improvement in the exposure by factors of 2 (sky maps) to 4 (spectra) over the previous measurements~\cite{j1713_proc}. This improvement allows spectral and morphology studies of unprecedented precision, leading to detailed, spatially dependent studies of the SNR. The angular resolution of better than $0.05^{\circ}$ allows one to perform a detailed investigation of the morphological differences between the TeV and X-ray emission, yielding remarkable results. For the first time in TeV gamma-ray astronomy, VHE data allow one to construct the maps of physical parameters such as magnetic field. The magnetic field map obtained for RX\,J1713.7-3946 (Fig.\,\ref{j1713_Bmap}) shows that this quantity is very variable across the remnant. A comparison of the TeV and X-ray radial profiles shows that the TeV shell is extended beyond the X-ray shell which may point either to particle escape or to a complicated configuration of the magnetic field~\cite{j1713_proc}. In any case these high-precision measurements show how the VHE gamma-ray astronomy can probe acceleration regions and open up new prospects for studies.

More than 300 SNRs are detected at radio frequencies, of which about 250 fall into the region of the HGPS, but only $\sim50$ are coincident with VHE sources. This underdetection of SNRs at TeV energies motivated an SNR population study~\cite{SNRpop_proc} providing flux upper limits for 124 sources. The study showed a clear correlation between VHE flux to radio flux ratio and source age. This kind of study might be very useful for future observations of SNR with the Cherenkov Telescope Array (CTA). 

\vspace*{-3ex}
\section*{\sc pulsars}
\vspace*{-1ex}
\indent\indent Pulsars are rapidly rotating and highly magnetised neutron stars created as a result of supernovae explosions. They are surrounded by a rotating magnetosphere and feature relativistic outflows. Pulsars emit pulsed emission at all wavelengths and although they were primarily detected at radio frequencies (with $\sim2500$ radio pulsars detected so far), most of their radiation is believed to be emitted at high energies via curvature radiation of charged particles (electrons and positrons) accelerated in the electromagnetic field of the pulsar. This is supported by a rapid increase of pulsar detections at GeV energies in recent years thanks to the new sensitive instruments \fermi\ and AGILE, with numbers reaching now more than 150 objects~\cite{2013ApJS..208...17A}. The energy spectra of most of the gamma-ray pulsars can be well described by an exponentially cut-off power law, $E^{-\Gamma}\exp\left[-\left(E/E_{\mathrm{cut}}\right)^b\right]$, with $b\leq1$ and cut-off energy $E_{\mathrm{cut}}$ typically between 1 and 10\,GeV~\cite{2013ApJS..208...17A}. A sub-exponential cut-off supports models of gamma-ray production in the outer magnetopshere, excluding a polar cap model for which a super-exponential cut-off ($b>1$) is expected. The extrapolation of pulsar spectra detected by \fermi\ to higher energies reveals a dramatic decrease of the gamma-ray flux beyond $10$\,GeV, which makes the detection of pulsars at energies $\gtrsim100$\,GeV with current ground-based instruments very unlikely. However, quite surprisingly, the first detection of the Crab pulsar above $25$\,GeV by MAGIC~\cite{2008Sci...322.1221A} with the flux consistent with the extrapolation of \fermi\ spectrum was followed by further detections of pulsed gamma-ray emission, first up to 250\,GeV by VERITAS~\cite{2011Sci...334...69V}, and later up to 400\,GeV by MAGIC~\cite{2012A&A...540A..69A}. The nature of this emission is still not understood with several explanations being suggested, such as inverse Compton upscattering of the magnetospheric X-ray emission by the pulsar wind electrons~\cite{2012Natur.482..507A} or the IC emission of secondary electrons in the outer magnetosphere~\cite{2012ApJ...754...33L,2006MNRAS.366.1310T}. Recently 320\,h of observations allowed MAGIC to extend the spectrum up to $\sim2$\,TeV~\cite{crabmagic_proc}, providing a further support for the IC models. However, VERITAS (with 194\,h of observations) did not confirm this result, revealing a firm detection of the pulsed emission only up to 400\,GeV~\cite{2015arXiv150807268N}.

One of the major science objectives for the new 28\,m H.E.S.S. telescope was to pursue the pulsar observation program, providing more information for the understanding of the nature of the pulsed VHE radiation. The principal source chosen for this purpose was the Vela pulsar, the brightest source in the high energy gamma-ray sky with a hint of pulsed emission above 20\,GeV observed using the \fermi\ data. The data was taken only with the 28\,m telescope, in the monoscopic way, providing a firm detection of the pulsed radiation in the energy band from $20$\,GeV  to 120\,GeV, establishing a second VHE pulsar~\cite{velapulsaar_proc}. 

\vspace*{-3ex}
\section*{\sc pulsar wind nebulae}
\vspace*{-1ex}
\indent\indent The electron-positron plasma ejected from energetic pulsars in the form of relativistic winds carries most of the rotational energy of the pulsars. The pulsar wind interacting with the ambient medium terminates at a standing shock where particles can be efficiently accelerated. Accelerated leptons can interact with the magnetic field and low-energy photon fields, generating non-thermal emission from radio frequencies to energies as high as $100$\,TeV. This results in the formation of a synchrotron nebula around the pulsar seen in radio to X-rays and more extended IC nebula at GeV and TeV energies.

Pulsar wind nebulae appear to be the most effective gamma-ray emitters in the Galaxy, forming the most numerous class of VHE gamma-ray Galactic objects. The list of 12 firmly identified VHE PWNe detected in the HGPS can be completed by 6 PWNe outside the HGPS and about one third of the 50 unidentified sources which are coincident with young powerful pulsars. TeV PWNe detected by \hess\ can naturally be divided into two classes based on their morphology, which in turn serves as an indication of the pulsar age. Young PWNe such as the Crab Nebula~\cite{2006A&A...457..899A}, G0.9$+$0.1~\cite{2005A&A...432L..25A}, G21.5$-$0.9~\cite{2008ICRC....2..823D}, etc., are generally detected as compact and unresolved objects. In such systems the TeV emission region is coincident with the associated young high spindown luminosity pulsar and is compatible with the X-ray emission region. Older PWNe, such as Vela\,X~\cite{2012A&A...548A..38A,2006A&A...448L..43A}, HESS\,J1825-137~\cite{2005A&A...442L..25A, 2006A&A...460..365A} and HESS\,J1303-631~\cite{2005A&A...439.1013A, 2012A&A...548A..46H}, show much more complicated morphologies, with the TeV emission regions much larger than the X-ray emission regions and pulsars significantly offset from the centre of the nebula. The larger size of the VHE PWN comparing to the X-ray one can be explained by synchrotron cooling of very energetic electrons. Very energetic electrons producing the X-ray emission via synchrotron radiation undergo strong radiative losses and lose their energy relatively fast.  At the same time, electrons need less energy to produce TeV emission via IC scattering. These electrons suffer less from radiative cooling and therefore can survive longer and in greater number. This scenario was supported by the detailed energy-dependent morphology studies in HESS\,J1825-137 and HESS\,J1303-631. Steepening of the spectrum with the distance from the pulsar detected in HESS\,J1825-137 (Fig.\,\ref{j1825}) clearly indicates the radiative cooling of electrons and gives an insight into the PWN evolution, allowing one to look into older epoches. Similarly, the energy-dependent morphology study of HESS\,J1303-631 showed that the emission region ``shrinks'' towards the position of the pulsar with the increase of the energy threshold (Fig.\,\ref{j1303}). 

The offset of the pulsar can be explained by the proper motion of the pulsar due to the initial kick obtained in the supernova explosion and/or by the destruction of a part of the nebula by the reverse shock of an SNR. The latter is believed to be the case for the PWN of Vela pulsar, Vela\,X. North-eastern and south-western sides of the Vela SNR are believed to be expanding into the media with different particle densities~\cite{2014A&A...561A.139S,2011A&A...525A.154S} which leads to faster formation of the reverse shock on the side with higher density~\cite{2001ApJ...563..806B}. Therefore it is possible that on one side the reverse shock has already reached the PWN while on the other side they still did not interact.

From the beginning of the VHE astronomy era, \hess\ detected many so-called ``dark'' sources. These are the sources detected only at TeV energies without counterparts at radio or X-ray energies. A majority of these sources are significantly extended, and a lot of them are coincident with energetic pulsars. It has been suggested recently that a substantial fraction of these ``dark accelerators'' might be the evolved PWNe~\cite{2009arXiv0906.2644D}. It was shown that the magnetic field in PWNe decreases with time, hence leading to the suppression of the synchrotron emission, while the IC emission increases with time until most of the pulsar spindown energy is transferred to the nebula. One of the best examples of previously ``dark'' emitters which were identified as PWNe is HESS\,J1303$-$631. Its identification as a PWN was based on the energy-dependent morphology which indicated the association with the pulsar PSR\,J1301$-$6305 and on the subsequent detection of the X-ray counterpart~\cite{2012A&A...548A..46H}. The source was subsequently detected at GeV energies by \fermi, exhibiting a similar morphology as at TeV energies with a larger emission region~\cite{2013ApJ...773...77A}. Recent dedicated radio observations with ATCA did not reveal any significant extended emission associated with the pulsar, but a shell-like structure, possibly an SNR, was detected in the field of view~\cite{2015arXiv150901427S}. In case it is an SNR, it might be the birth place of the pulsar. 

\vspace*{1ex}
\begin{figurehere}
\centering
\epsfig{file=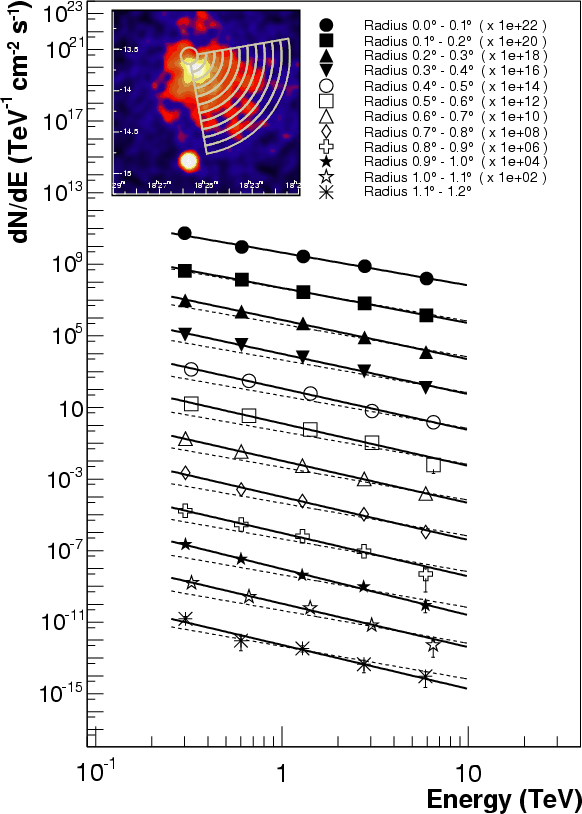,width=\linewidth}
\caption{Energy spectra of HESS\,J1825-137 in radial bins, reflecting the steepening of the spectrum with the distance from the pulsar. The HESS excess map is shown in the inset. The wedges show the radial regions with radii in steps of $0.1^{\circ}$ at which the energy spectra were determined. The innermost region is centred on the pulsar PSR J1826-1334. The differential energy spectra for the regions illustrated in the inset are scaled by powers of 10 for the purposes of clarity. The spectrum for the analysis at the pulsar position is shown with the dashed line as a reference along with the other spectra. The figure is taken from~\cite{2006A&A...460..365A}.}\label{j1825}
\end{figurehere}

A large sample of identified TeV PWNe and a comparable amount of PWN candidates detected only in the TeV range motivated the population study of these objects~\cite{2013arXiv1307.7905K, pwnpop}. For the first time the HGPS also allows for the extraction of flux upper limits from the regions around pulsars without detected TeV emission. All this information allows for a systematic investigation of the evolution of parameters such as luminosity and extension over $\sim10^5$ years after the birth of the pulsar. Population studies reveal some trends in the evolution of PWNe, such as a decrease of spin-down luminsity with age, expansion of PWNe with time and the fading of old PWNe, but there are also sources which exhibit large variations from the average behaviour, which are likely due to the diversity of ambient media and intrinsic initial conditions.

\vspace*{1ex}
\begin{figurehere}
\centering
\epsfig{file=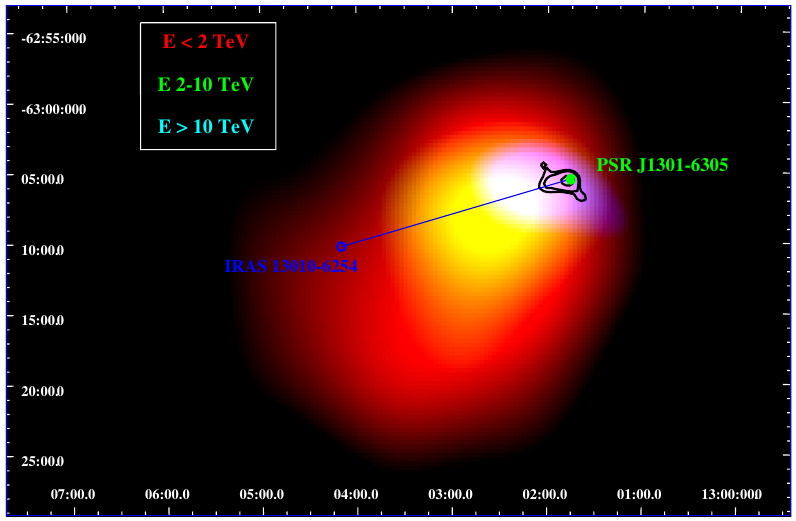}
\caption{Energy mosaic of HESS\,J1303$-$631. The horizontal axis is the Right Ascension and the vertical axis is the Declination in J2000.0 coordinates. Red, green and blue colours indicate different energy ranges: $E_1 = (0.84 - 2)$\,TeV, $E_2 = (2 - 10)$\,TeV and $E_3 > 10$\,TeV, respectively. The highest energy photons originate nearest to the pulsar, PSR J1301-6305 (marked by the green dot). The visible red corresponds roughly to the $10\sigma$ significance contour of the entire source. XMM-Newton X-ray contours are shown in black. The figure is taken from~\cite{2012A&A...548A..46H}.}\label{j1303}
\end{figurehere}

\vspace*{-3ex}
\section*{\sc gamma-ray binaries}
\vspace*{-1ex} 
\indent\indent Gamma-ray binaries comprise a relatively small class of VHE sources consisting only of 5 objects. Binary systems are variable sources consisting of a massive star and a compact object such as a black hole or a pulsar. The TeV emission in these systems is believed to originate from the interaction between the two objects, either in the accretion-powered jet, or in the shock between the pulsar wind and the stellar wind.

The five binaries detected at TeV energies are PSR\,B1259$-$63/LS 2883~\cite{2009A&A...507..389A,2005A&A...442....1A,  2013A&A...551A..94H}, LS\,5039, HESS\,J0632$+$057~\cite{2007A&A...469L...1A, 2014ApJ...780..168A}, LSI\,$+$61\,303~\cite{2008ApJ...679.1427A,2006Sci...312.1771A}, and HESS\,J1018$-$589 (1FGL\,1018.6$-$5856)~\cite{2012A&A...541A...5H, 2015A&A...577A.131H}. HESS\,J0632$+$057 is the first binary primarily discovered at TeV energies~\cite{2009ApJ...690L.101H} and the only one which can be observed in both the northern and southern sky. HESS\,J1018$-$589 is a new member of the class of the TeV gamma-ray binaries. Recently, re-observations of this point-like source, coincident with the high-energy binary 1FGL\,1018.6$-$5856 detected by \fermi\ \cite{2010ApJS..188..405A}, revealed its variability at TeV energies~\cite{2015A&A...577A.131H}. The consistency of the TeV light curve of the source with the GeV and X-ray light curves (Fig.\,\ref{j1018lc}) confirms the association of HESS\,J1018$-$589 with 1FGL\,1018.6$-$5856. 

\vspace*{1ex}
\begin{figurehere}
\centering
\epsfig{file=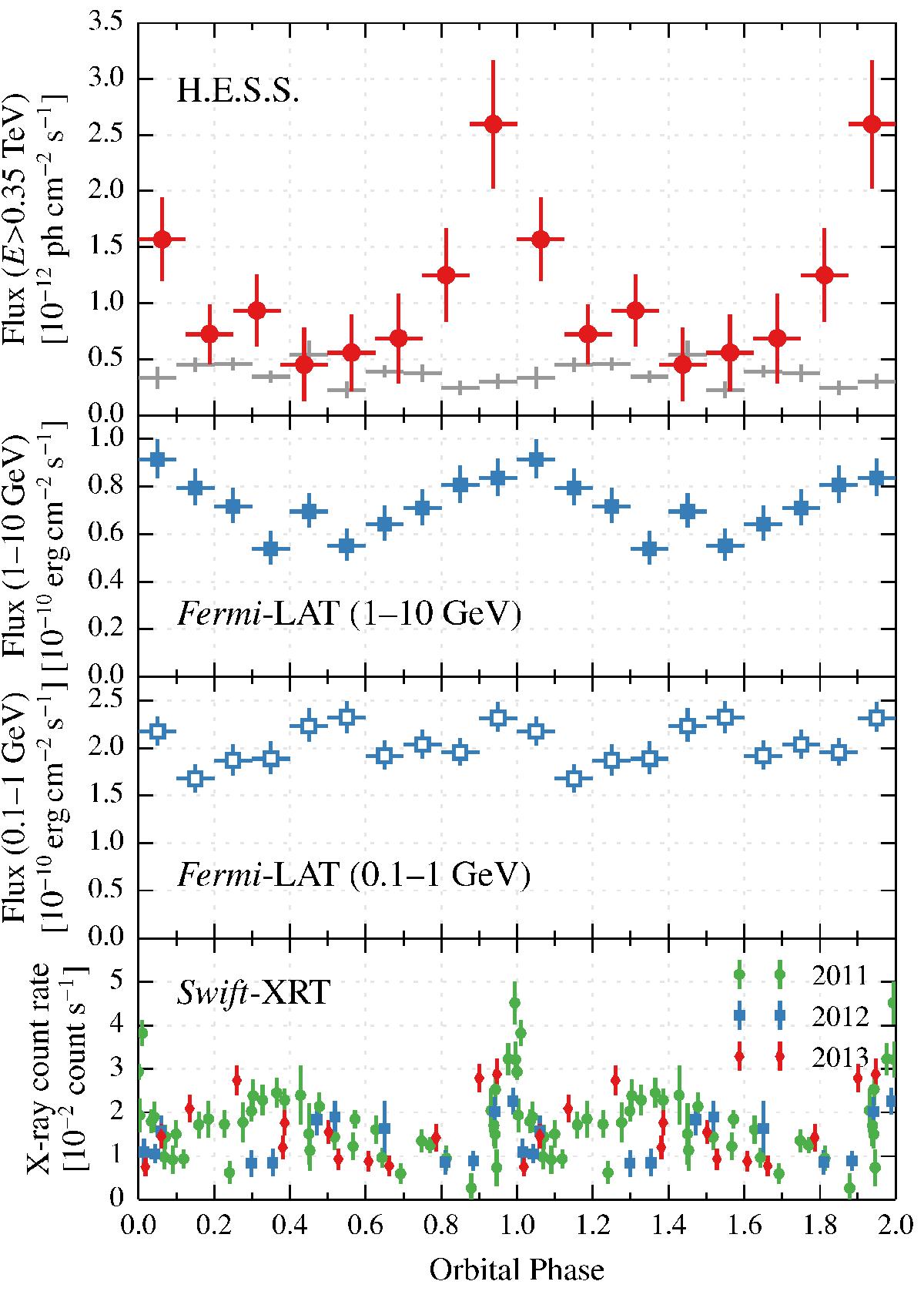, width=\linewidth}
\caption{Gamma-ray and X-ray fluxes of 1FGL\,J1018.6$-$5856 folded with the orbital period of $P = 16.58$\,d. Two orbits are shown for clarity. Top: VHE integral flux above 0.35\,TeV measured by H.E.S.S. (red circles). For comparison, a scaled light curve from the nearby bright source HESS\,J1023$-$589 is shown in grey. Middle top and middle bottom: \fermi\ light curve between 1 and 10 GeV (solid blue squares) and between 0.1 and 1 GeV (open blue squares). Bottom: X-ray $0.3-10$\,keV count-rate light curve from Swift-XRT observations in 2011 (green), 2012 (blue), and 2013 (red). The figure is taken from~\cite{2015A&A...577A.131H}.}\label{j1018lc}
\end{figurehere}
\vspace*{1ex}

\begin{figure}[!t]
\centering
\begin{minipage}{.98\linewidth}
\centering
\epsfig{file=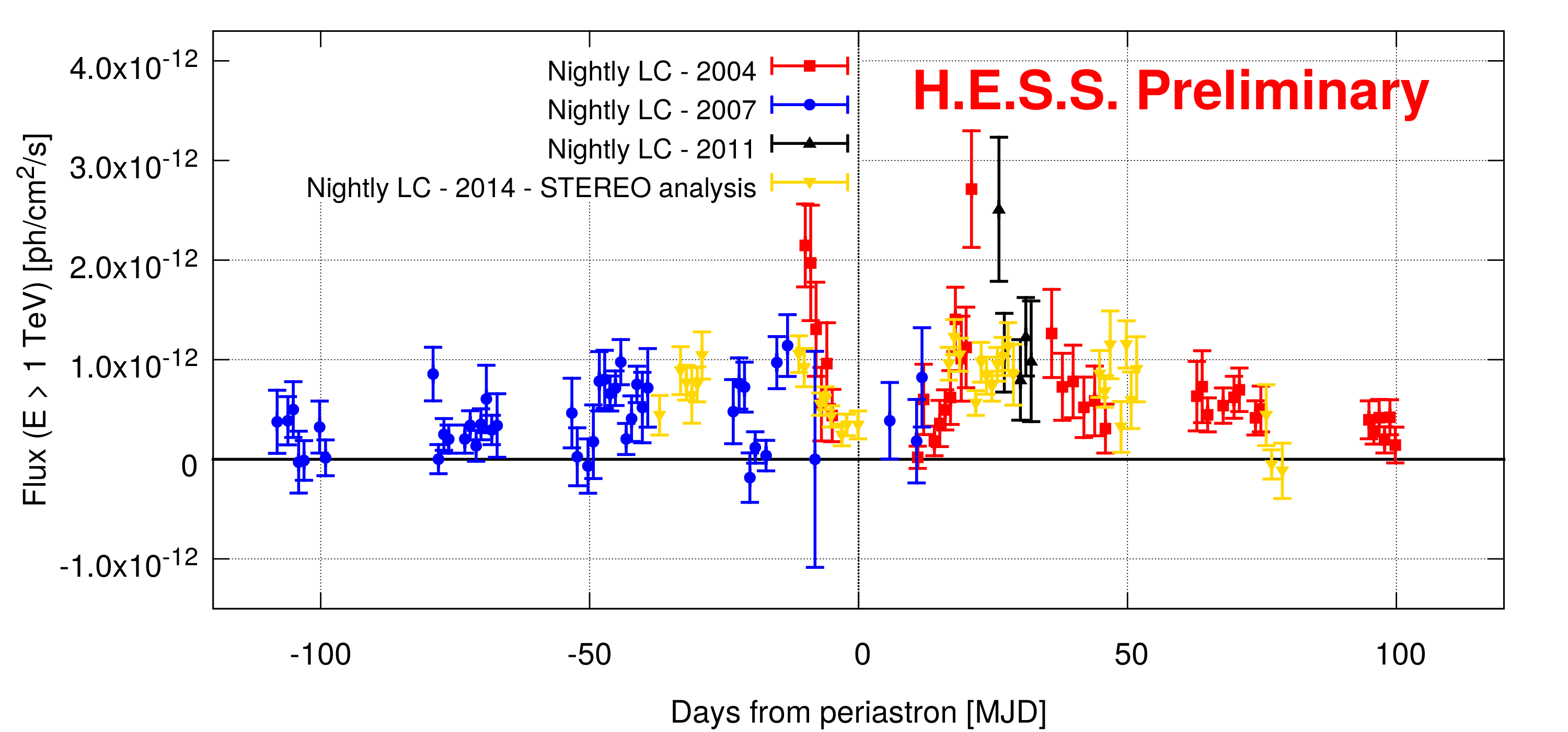,width=\linewidth}
\caption{A VHE light curve of PSR\,B1259-63/LS\,2883 combining data from four periastron passages in 2004 (red points), 2007 (blue points), 2011 (black points), and 2014 (yellow points). Data points show the nightly flux above 1\,TeV from about 100 days before to 100 days after the periastron passage. The figure is taken from~\cite{2015arXiv150903090R}.}\label{psrb1259lc}
\end{minipage}
\end{figure}

The only TeV binary for which the nature of the compact object is well known is PSR\,B1259$-$63/LS\,2883. It consists of a pulsar orbiting a Be star in a very eccentric orbit ($e=0.87$) with a period of $3.4$~years. Since the start of \hess\ operation, the periastron passage in the system has occurred four times, in 2004~\cite{2005A&A...442....1A}, 2007~\cite{2009A&A...507..389A}, 2010~\cite{2013A&A...551A..94H}, and 2014~\cite{2015arXiv150903090R}, thoroughly observed by \hess each time. In 2014, the source was visible for the first time directly at the periastron crossing and also for the first time it was possible to observe the source both before and after the periastron passage. This allowed for the confirmation of the light curve shape obtained from the combined observation of three previous periastron passages, showing that it does not change from orbit to orbit. VHE observations show no emission far from periastron and a complex light curve at the periastron passage exhibiting two peaks, before and after periastron (Fig.\,\ref{psrb1259lc}). The TeV flux variability has a similar shape as the X-ray and radio emission, featuring peaks at the same orbital phases. The nature of the VHE emission can be explained as IC radiation within the pulsar-wind stellar-wind scenario. The shape of the light curve is not well understood yet, but it is believed that pre- and post-periastron peaks are related to the location of the equatorial circumstellar disk of the Be star which the pulsar crosses twice each orbit.

At GeV energies, however, {\fontsize{10}{11}\selectfont PSR\,B1259$-$63/LS\,2883} shows a completely different behaviour, displaying a remarkable post-periastron flare which is time-shifted with respect to the post-periastron peak at other wavebands~\cite{2011ApJ...736L..11A,2015arXiv150801339C, 2011ApJ...736L..10T, 2015ApJ...798L..26T}. First detected around the 2010 periastron passage, the flare then re-appeared with a slightly lower flux at the same orbital phase during the 2014 periastron passage, revealing a periodic behaviour of this phenomenon. Apart from the flare, GeV observations around 2010 periastron passage also showed a faint detection close to periastron~\cite{2011ApJ...736L..11A, 2011ApJ...736L..10T} which, however, was not confirmed during the 2014 passage~\cite{2015arXiv150801339C,2015ApJ...798L..26T}. The nature of the flare is still not understood. Several explanations for this have been suggested (see e.\,g.~\cite{2008MNRAS.387...63B, 2010A&A...516A..18D,2013A&A...557A.127D,2012ApJ...752L..17K, 2012ApJ...753..127K}), but each has its limitations. 

In 2014, the source was for the first time observed in the \hess II phase, exploiting the new 28\,m telescope. The data collected with \hess II allowed for the extention of the spectrum down to $200$\,GeV, resulting in a spectrum harder than during previous periastrons~\cite{2015arXiv150903090R}. New observations also revealed a rather high flux from the source 50 days after periastron during the period overlaping with the GeV flare. Although these results are still preliminary and careful data analysis is still ongoing, this new information may inspire new efforts towards a better understanding of the unexpected GeV flare.

\vspace*{-3ex}
\section*{\sc summary}
\vspace*{-1ex}
\indent\indent This paper discusses only those results from the Galactic VHE gamma-ray astronomy that are the most recent and the most interesting according to the admittedly biased opinion of the author. Other VHE gamma-ray astronomy discoveries go far beyond our own Galaxy -- e.\,g. detecting galactic sources in the Large Magelanic Cloud and observing active galactic nuclei up to $z\sim1$. Among the other targets of the gamma-ray astronomy is the search for the dark matter annihilation. The remarkable results achieved over the last decade in the field of the VHE gamma-ray astronomy exceeded all expectations, providing a massive boost to the development of theoretical studies in the fields of particle acceleration and radiation processes. The importance of gamma-ray astronomy was highly recognised by the scientific community, resulting in the development of new projects, such as CTA and HAWC, which will further increase the sensitivity and resolution of the gamma-ray observations thereby providing deep insights into a number of physical problems.

\end{multicols}

\end{document}